\documentclass[a4paper
,twocolumn,10pt
]{article}
\usepackage{epsfig} 
\usepackage{times}
\usepackage{graphicx}
\usepackage{cite}
\usepackage{citesupernumber}
\usepackage{naturefem}
\usepackage{nature}
\usepackage{amsmath}
\usepackage{color}
\ifx\pdfoutput\undefined
   \DeclareGraphicsExtensions{.epsi,.eps}
\else
   \DeclareGraphicsExtensions{.pdf}
\fi
\newcommand{\unit}[1]{\ensuremath{\operatorname{#1}}}
\newcommand{\mycite}[1]{\cite{#1}}
\newcommand{\new}[1]{%\textcolor{blue}
{#1}}

\begin{document}

\noindent{\Large\bf Cavity cooling of a single atom}\\[3mm]
\noindent {\small{\bf P. Maunz, T. Puppe, I. Schuster, N. Syassen, P.W.H. Pinkse \& G.~Rempe}\\
\noindent {\it Max-Planck-Institut f\"ur Quantenoptik,\\
 Hans-Kopfermann-Str. 1, D-85748 Garching, Germany}\\[2mm]}

\noindent \textbf{%
  All conventional methods to laser-cool atoms rely on repeated cycles of
  optical pumping and spontaneous emission of a photon by the atom.
  Spontaneous emission \new{in a random direction} is the dissipative
  mechanism required to remove entropy from the atom. However, alternative
  cooling methods have been proposed\mycite{HorakPRL97,VuleticPRL00} for a
  single atom strongly coupled to a high-finesse cavity; the role of
  spontaneous emission is replaced by the escape of a photon from the cavity.
  Application of such cooling schemes would improve the performance of atom
  cavity systems for quantum information processing\mycite{Kuhn02,Monroe02}.
  Furthermore, as cavity cooling does not rely on spontaneous emission, it can
  be applied to systems that cannot be laser-cooled by conventional methods;
  these include molecules\mycite{VuleticPRL00} (which do not have a closed
  transition) and collective excitations of Bose
  condensates\mycite{Horak2001}, which are destroyed by randomly directed
  recoil kicks. Here we demonstrate cavity cooling of single rubidium atoms
  stored in an intracavity dipole trap. The cooling mechanism results in
  extended storage times and improved localization of atoms. We estimate that
  the observed cooling rate is at least five times larger than that produced
  by free-space cooling methods, for comparable excitation of the atom.  }

The basic idea behind cavity cooling can be understood from a simple classical
picture based on the notion of a refractive index. Consider a standing-wave
optical cavity resonantly excited by a weak probe laser blue detuned from the
atomic resonance. For strong atom-cavity coupling, even one atom can
significantly influence the optical path length between the cavity mirrors.
Consequently, the intracavity intensity is strongly affected by the
atom\mycite{MabuchiOL96,MunstermannOptCom99,Sauer03}. For example, at a node
of the standing wave the atom is not coupled to the cavity, thus the
intracavity intensity is large. An atom at an antinode, in contrast, shifts
the cavity to a higher frequency because the atom's refractive index is
smaller than unity above its resonance. This tunes the cavity out of resonance
from the probe laser and leads to a small intracavity intensity. However, in a
high-finesse cavity the intensity cannot drop instantaneously when the atom
moves away from a node. Instead, the blue-shift of the cavity frequency leads
to an increase of the energy stored in the field. The photons finally escaping
from the cavity are therefore blue-shifted from the photons of the probe
laser. This occurs at the expense of the atom's kinetic energy. The reverse
effect, namely the acceleration of an atom approaching an antinode, is much
smaller as here the cavity is initially out of resonance with the probe laser
and consequently the intracavity intensity is small.

Note that the cooling process does not require atomic excitation. Indeed, the
atomic excitation is low at all times as the atom is not coupled to the light
at a node while the intracavity intensity is very low for an atom near an
antinode. \new{It follows that the lowest attainable temperature is not
  limited by the atomic linewidth as for free-space Doppler cooling but by the
  linewidth of the cavity, which can be much smaller. Therefore temperatures
  below the Doppler limit can be reached\mycite{HechenblaiknerPRA98}. An upper
  limit on the velocity of the atom to be cooled is given by the requirement
  that the atom must not move farther than about one-quarter of a wavelength
  during the lifetime of a photon in the cavity. In our experiment, this
  corresponds to a velocity of about $3\unit{m/s}$.} We emphasise that cavity
cooling is applied to a single two-level atom and differs from the mechanical
effects observed for an atomic ensemble\mycite{ChanPRL03,Nargorny03,Kruse03}.
\new{A description of cavity cooling in terms of dressed-states pictures of
  the strongly coupled atom-cavity system can be found
  elsewhere\mycite{HechenblaiknerPRA98}.}
  
A treatment of cavity cooling combined with trapping by means of an auxiliary
far-red detuned dipole laser \new{ is quantitatively different.  It} can be
achieved by including the dynamic Stark shift of the atomic ground and excited
states into the Hamiltonian. The dynamic Stark shift renders the atomic
resonance frequency position dependent, making it larger for an atom at an
antinode. This effect even enhances the cooling force by effectively
increasing the refractive index variations for a moving atom.  The combined
system can be investigated numerically\mycite{vanEnkPRA01}.  Moreover, in the
limit of low atomic excitation analytic expressions for all relevant forces
including the cooling force can be derived, thereby extending previous
calculations\mycite{HorakPRL97,HechenblaiknerPRA98}. The obtained expressions
are lengthy but valuable for parameter optimisation and straightforward
trajectory calculation.

%%%%%%%%%%%%%%%%%%% figure 1 %%%%%%%%%%%%%%%%%%%%%%%%%%%%%%%%%%%
\begin{figure}[hb]
\begin{center}
\epsfig{file=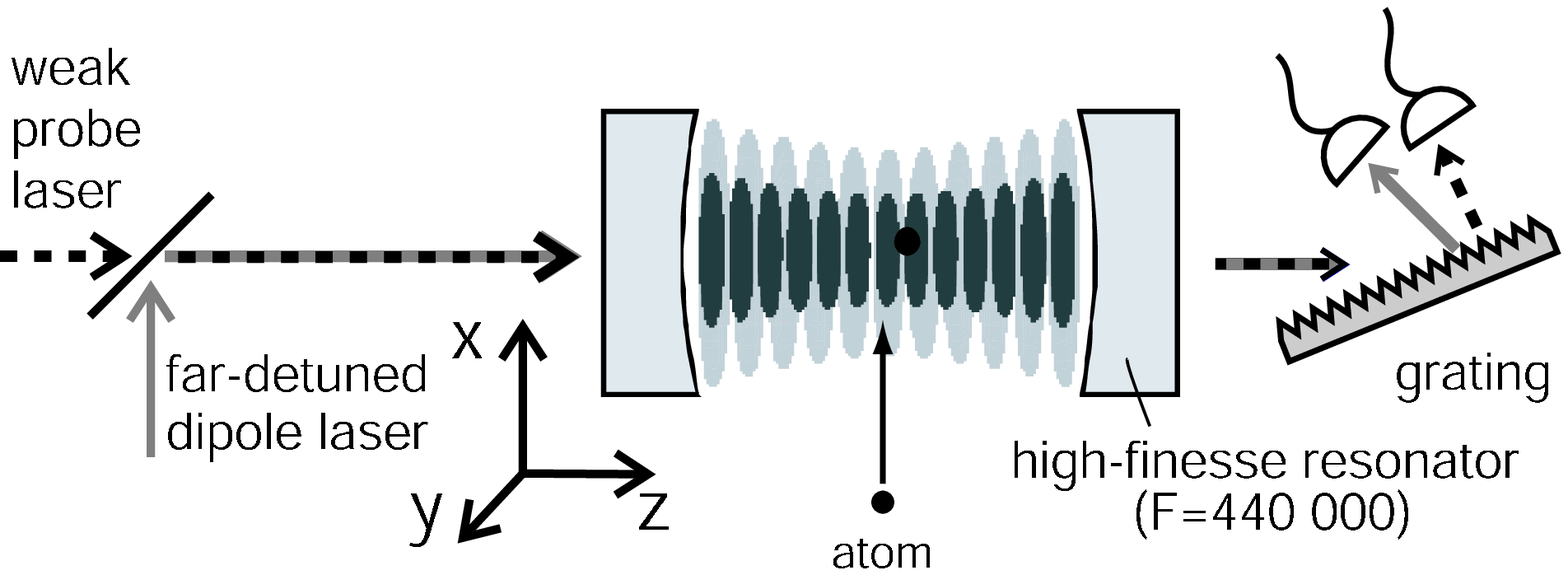,width=0.95\columnwidth}
\caption{\small\textbf{Experimental set-up.} The high-finesse cavity $({\cal
    F}=4.4\times 10^5)$ is excited by a weak near-resonant probe field and a
  strong far-red detuned dipole field.  $^{85}$Rb atoms are injected from
  below. Behind the cavity, the two light fields are separated by a grating.
  The probe light is further passed through a narrow-band interference filter
  before being directed onto a single-photon counting module. For this set-up, a
  quantum efficiency of 32\% is achieved for the detection of probe light
  transmitted through the cavity while the dipole light is attenuated by more
  than $70\unit{dB}$.  The dipole light is also used to stabilize the cavity
  length with a radio frequency sideband technique. It is generated by a grating- and
  current-stabilized diode laser with a linewidth of 20 kHz
  r.m.s.}\label{system}
\end{center}
\end{figure}
%%%%%%%%%%%%%%%%%%% figure 1 %%%%%%%%%%%%%%%%%%%%%%%%%%%%%%%%%%%

%%%%%%%%%%%%%%%%%%%%%% The experimental setup, cavity  %%%%%%%%%%%%%%%%%%%%%%%%%
In our experiment, see Fig.~\ref{system}, the cavity has a finesse ${\cal
  F}=4.4\times 10^5$ and a length $l=120\unit{\mu m}$. The cavity field has a
decay rate $\kappa/2\pi=1.4\unit{MHz}$. The wavelength difference between
neighbouring longitudinal $\text{TEM}_{00}$ modes is about $2.5\unit{nm}$.
Single laser-cooled $^{85}$Rb atoms are injected into the
cavity\mycite{MunstermannOptCom99} with a velocity smaller than 10 cm/s. The
single-photon coupling constant is $g/2\pi=16\unit{MHz}$ for the
$5^2S_{1/2}F=3 \leftrightarrow 5^2P_{3/2}F=4$ transition with dipole decay
rate $\gamma/2\pi=3\unit{MHz}$. A weak, near-resonant probe laser at
$780.2\unit{nm}$ is used to observe and cool the atom. A strong, far-detuned
dipole laser at $785.3\unit{nm}$ serves to trap the atom.  The detunings of
the probe laser with respect to the cavity, $\Delta_c=0$, and the atom,
$\Delta_a/2\pi=35\unit{MHz}$, are chosen to compromise between ideal detunings
for detection and good cooling conditions while maintaining a very low
excitation of the atom at any moment of the experiment. In fact, the presence
of an atom at an antinode reduces the transmission of the probe light by a
factor of 100, allowing its detection and manipulation with a high
signal-to-noise ratio and a high
bandwidth\mycite{HoodScience00,PinkseNature00,MunstermannPRL99,FischerPRL02}.

Experiments are performed only with atoms located in the central region of the
cavity, where the nodes and antinodes of the probe and dipole fields coincide.
This is accomplished by turning on the dipole laser before injecting the atom
into the cavity. In this case, a $400\unit{\mu K}$ deep dipole potential
guides the arriving atom into the high-intensity region.  Only if this region
is also a region of strong coupling, the presence of an atom manifests itself
as a pronounced dip in the cavity transmission as monitored with the probe
light.  Atoms entering the cavity at an axial position where the two standing
waves are out-of-phase are confined to nodes of the probe field and, hence,
are invisible to the probe laser. If the transmission drops below 9\%, the
intensity of the dipole light is increased to generate a trap depth of about
$1.5\unit{mK}$. This compensates for the radial kinetic energy of the atom and
enables to catch the atom in an otherwise conservative potential. More than
95\% of the detected atoms are captured in the dipole trap.

%%%%%%%%% Lifetime of the bare dipole trap
The average storage time of the atom in the dark intracavity trap is measured
by turning off the probe light for an adjustable time interval, \new{$\Delta
  t$, after the atom is captured. As a function of the dark time, $\Delta t$,
  the fraction of atoms still trapped drops exponentially with a decay
  constant of} $18\unit{ms}$, \new{defining the storage time of the atoms in
  the trap.}  The theoretical limits imposed by light scattering
($85\unit{s}$) and genuine cavity QED dipole fluctuations
($200\unit{ms}$)\mycite{HorakPRL97} cannot explain this rather short
time\mycite{YePRL99,McKeeverPRL03}. Instead the observed loss of the atom is
attributed to parametric heating due to fluctuations of the intracavity
intensity mainly caused by frequency fluctuations of the dipole laser. This
technical noise critically depends on the tuning of the laser stabilization.
Indeed, the storage time in the dark dipole trap could be increased from
$18\unit{ms}$ to $31\unit{ms}$ by improving the frequency stability of the
dipole laser. As this stability and other sources of noise are hard to
control, a concurrent measurement of the storage time of the dark dipole trap
serves as a reference in each of the following experiments. Note that the
axial trap frequency is about 100 times higher than the radial trap frequency.
As parametric heating is proportional to the square of the trap
frequency\mycite{SavardPRA97} the atom is mainly heated in axial direction.
Since axial and radial motion are only weakly coupled, the heated atom usually
escapes the antinode of the standing wave dipole trap along the axis, thereby
hitting one of the mirrors.  This conjecture is supported by numerical
simulations of the experiment, as further discussed below.

%%%%%%%%%%%%%%%%%%% figure 2 %%%%%%%%%%%%%%%%%%%%%%%%%%%%%%%%%%%
\begin{figure}[bhtp]
\begin{center}
\epsfig{file=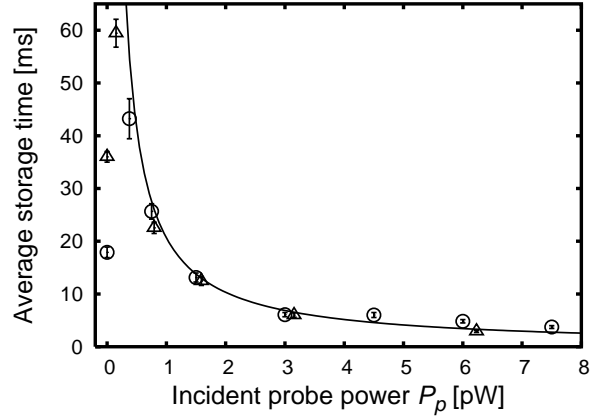,width=0.95\columnwidth}%
\caption{\small\textbf{Storage time.} The storage time of a
  trapped atom as a function of probe power. \new{Two sets of data taken
    before (circles) and after (triangles) improving the laser stabilization
    are displayed. The corresponding storage times in the dark trap are
    $18\unit{ms}$ and $36\unit{ms}$, respectively. For an incident probe power
    exceeding about $0.5\unit{pW}$ the storage time is limited by radial
    escape due to spontaneous emission. In this range the measured storage
    time (in ms) can be approximated by $20/P_p$ (where $P_p$ is in units of
    pW; solid line).  The storage time of $18\unit{ms}$ obtained for the dark
    trap increases by more than a factor of two by applying $0.37\unit{pW}$
    probe light. After the stabilization of the dipole laser was improved,
    $P_p=0.11\unit{pW}$ increased the storage time from $36\unit{ms}$ to
    $60\unit{ms}$. The mentioned probe powers correspond to an average
    intracavity photon number of $0.005$ and $0.0015$ for $0.37\unit{pW}$ and
    $0.11\unit{pW}$, respectively.} }
\label{StorageOverProbepower}
\end{center}
\end{figure}
%%%%%%%%%%%%%%%%%%% figure 4 %%%%%%%%%%%%%%%%%%%%%%%%%%%%%%%%%%%
%
%%%%%%%%%%%%%%%%%%%%%% Cooling force %%%%%%%%%%%%%%%%%%%%%%%%%%%
To demonstrate that cavity cooling can be used to compensate for the axial
heating of the dipole trap, the probe beam is not switched off completely
after capturing an atom. Fig.~\ref{StorageOverProbepower} shows the storage
time as a function of the incident probe power, $P_p$.  For high probe power
the storage time is reduced as compared to the dark trap.  However, the
storage time increases with decreasing power. This effect is attributed to the
reduction of spontaneous emission which heats the atom in all directions and
which cannot be compensated radially since cavity cooling acts mainly axially.
As the atomic transition is still far from being saturated, the radial heating
is proportional to the probe power even for the highest considered level of
$P_p=7.5\unit{pW}$. Hence, as long as the probe power is large enough to
compensate for axial heating, the storage time $\tau$ is limited by radial
loss and $\tau\propto P_p^{-1}$ (solid line). Consequently, the atom is
expected to leave the cavity axially for near-zero probe power, while in case
of higher power radial losses should dominate. This is confirmed by a Monte
Carlo simulation of a point-like atom moving in the trap under the influence of
the forces and momentum diffusion calculated analytically, while parametric
heating from the dipole trap is implemented by a randomly changing potential
depth. The storage times evaluated from the simulation agree well with the
experiment. Moreover it can be concluded that for probe powers below
$0.1\unit{pW}$ more than $90\%$ of the atoms leave the cavity by hitting a
mirror, while for higher probe powers $90\%$ of the atoms leave radially.

We emphasise that, in contrast to Doppler cooling, cavity cooling extends the
storage time for a probe field which is blue detuned from the atom. If the
detuning is changed from blue to red by adjusting the atom-cavity detuning
while keeping the dipole power constant and the probe laser resonant with the
cavity, the average storage time decreases and drops below the storage time of
the dark trap. This clearly demonstrates that the extension of the storage
time cannot be attributed to Doppler cooling.

%%%% cooling
%%%%%%%%%% figure 3 %%%%%%%%%%%%%%
\begin{figure}[htbp]
\begin{center}
\epsfig{file=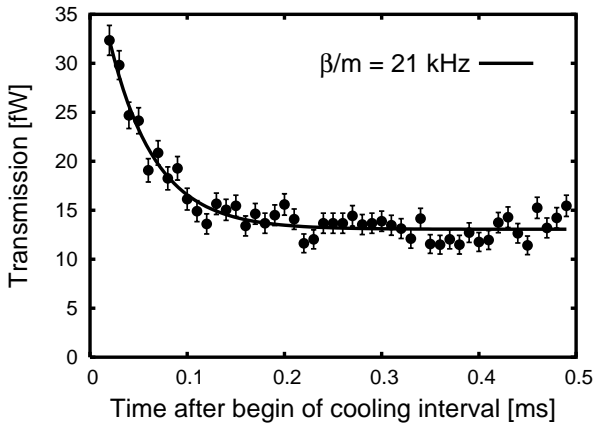,width=0.95\columnwidth}%
\caption{\small\textbf{Cooling-induced localization.} Average transmission
  during cooling intervals \new{of length $0.5\unit{ms}$ after heating the atom for
    $0.1\unit{ms}$. An incident power of $P_p=2.25\unit{pW}$ is chosen for a
    good signal to noise ratio. Without an atom, the cavity transmission on
    resonance is $300\unit{fW}$.} The atoms are cooled during the first
  $0.1\unit{ms}$. This leads to a stronger coupling to the cavity mode and,
  hence, to a smaller transmission. A cooling rate of $\beta/m=21\unit{kHz}$
  is estimated from an exponential fit.  \new{Radial heating occurs on a much
    longer timescale and is not visible here.} }
\label{CoolingTransmission}
\end{center}
\end{figure}
%%%%%%%%%% figure 3 %%%%%%%%%%%%%%
%
We now demonstrate cooling by directly observing the reduction of the kinetic
energy of the atom. The experiment employs alternating heating and cooling
intervals at a constant probe power of $P_p=2.25\unit{pW}$. Axial heating is
achieved by deliberately tuning the probe laser for a time interval of
$100\unit{\mu s}$ to a frequency $9\unit{MHz}$ above the cavity
\new{resonance} ($\Delta_c/2\pi=9\unit{MHz}$), \new{where strong heating of
  the atom is expected from the theoretical analysis.} In the following
$500\unit{\mu s}$ long cooling interval the probe laser is switched back to
the cavity resonance ($\Delta_c=0$).  Fig.~\ref{CoolingTransmission} shows the
cavity transmission averaged over many cooling intervals, \new{with an atom
  present in the cavity.} The transmission drops by more than a factor of two
during the first $100\unit{\mu s}$.  This drop is a clear signature for the
increasing atom-cavity coupling, and hence, a better localization of the atom
at the antinode.

The exponential change of the transmission observed for short times allows to
obtain an estimate of the mean cooling rate, $\beta/m=21\unit{kHz}$, where
$\beta$ is the friction coefficient and $m$ the atomic mass. \new{This result
  is in good agreement with the Monte Carlo simulations, which show an
  exponential relaxation with a timescale of about $50\unit{\mu s}$ for the
  increase of the atomic localization as well as for the decrease of the
  transmitted power.} To compare this rate of cavity cooling with free-space
cooling rates of a two-level atom for a given rate of spontaneous emission
events, knowledge about the atomic excitation is required. Here, an upper
limit can be obtained by attributing the storage time of $9\unit{ms}$ (measured
for a probe power of $2.25\unit{pW}$; Fig.~\ref{StorageOverProbepower})
solely to radial heating from spontaneous emission. To leave the trap, an atom
must have gained about $1\unit{mK}$ of kinetic energy. This limits the atomic
excitation to below $2.5\%$. At this excitation, free-space Sisyphus
cooling\mycite{Aspect86,Dalibard85} \new{ of a two-level atom in a
  blue-detuned standing wave} would achieve $\beta_{\text{S}}/m=4\unit{kHz}$
while Doppler cooling would have $\beta_{\text{D}}/m=1.5\unit{kHz}$, \new{both
  for optimal detuning}. Thus introducing the cavity increases the cooling
rate by at least a factor of 5 for constant atomic excitation.

%%%%%%%%%% figure 4 %%%%%%%%%%%%%%
\begin{figure}[htbp]
\begin{center}
\epsfig{file=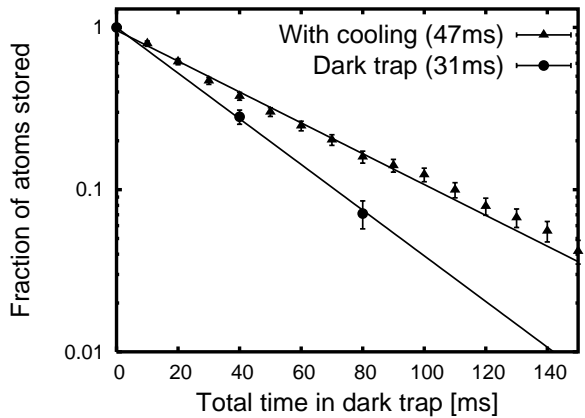,width=0.95\columnwidth}%
\caption{\small\textbf{Cavity cooling.} The fraction of atoms stored in the trap as 
  a function of time after the dipole trap is switched on. The dark dipole
  trap has an average storage time of $31\unit{ms}$ (circles), as obtained
  from an exponential fit. If $100\unit{\mu s}$ long cooling intervals are
  applied every $2\unit{ms}$, the average storage time without counting the
  cooling intervals is extended to $47\unit{ms}$ (triangles). \new{The storage
    time in the dark trap varies slightly from day to day and is therefore
    measured concurrently. As in several detailed measurements (not shown) the
    fraction of atoms found in the dark trap as a function of time was found
    to be very well described by an exponential decay, three data points
    (including the one at zero trapping time) are sufficient to obtain the
    storage time in the dark trap.} }
\label{CoolingDecay}
\end{center}
\end{figure}
%%%%%%%%%% figure 4 %%%%%%%%%%%%%%

Cooling \new{down} an atom in the trap can also be demonstrated without the
additional heating. For this purpose the atom is repeatedly left in the dark
dipole trap by switching off the probe light during $2\unit{ms}$ long time
intervals.  These intervals are short compared to the average trapping time in
the dark trap of $31\unit{ms}$, \new{obtained after improving the frequency
  stability of the dipole laser.} In each of the dark intervals the atom
experiences parametric heating. Between the dark intervals, $100\unit{\mu s}$
long cooling intervals are applied using a probe power $P_p=1.5\unit{pW}$ on
resonance with the cavity ($\Delta_c=0$). The transmission of the probe light
is also used to determine whether the atom is still present.  The average time
an atom is stored in the trap under these conditions can be calculated by
adding up all the $2\unit{ms}$ long dark intervals after which the atom is
still found in the trap, but omitting the cooling intervals.  The result is
shown in Fig.~\ref{CoolingDecay}: \new{although} the short cooling intervals
\new{have a duty cycle of only 5\% they} increase the average trapping time by
more than 50\%. Obviously, heating the atom out of the trap requires more time
in the presence of cooling. Therefore the kinetic energy of the atom was
reduced during the cooling interval.

%%%%%%%%%%%%%%%%%%%%%% conclusion and outlook %%%%%%%%%%%%%%%%%%%%%%%%%
In conclusion, strong coupling of a two-level atom to the cavity field was
used to cool single atoms stored in an intracavity dipole trap. The storage
time in the trap has been increased by a factor of two by exploiting the
cooling force caused by a near-resonant cavity field with an average photon
number of only $0.005$. In contrast to free-space laser cooling techniques,
this cooling force acts mainly by exciting the cavity part of the coupled
atom-cavity system. Thus strong cooling forces can be achieved while keeping
the atomic excitation low. An estimate of the strength of the cooling force
has shown to exceed the force expected for free-space Sisyphus cooling and
Doppler cooling at comparable atomic excitation by at least a factor of 5 and
14, respectively. Avoiding excitation could serve as a basis for cooling of
molecules or collective excitations of a Bose condensate\mycite{Horak2001}.
Another application might be to cool the motion of an atom with a stored
quantum bit\mycite{Griessner03}. If the two states forming the qubit have identical
coupling to the cavity, the new cooling scheme would not disturb the
superposition state. This advantage is not shared by any other cooling method.

%%%%%%%%%%%%%%%%%%%%%%%%%%%%% Acknowledgements %%%%%%%%%%%%%%%%%%%%%%%%%%%%%%%
%%%%%%%%%%%%%%%%%%%%%%%%%%%%%%% References %%%%%%%%%%%%%%%%%%%%%%%%%%%%%

\noindent\textbf{Correspondence} and requests for materials should be addressed to G.\,R.\
(e-mail: gerhard.rempe@mpq.mpg.de).


\begin{thebibliography}{}
\fontsize{8}{10pt}\selectfont 
\itemsep=0mm
\bibitem{HorakPRL97} P.~Horak, G.~Hechenblaikner, K.M.~Gheri, H.~Stecher, and
  H.~Ritsch, Cavity-Induced Atom Cooling in the Strong Coupling Regime.
  \textit{Phys.~Rev.~Lett.}~{\bf 79}, 4974-4977 (1997).

\bibitem{VuleticPRL00} V. Vuleti{$\rm \acute{c}$} and S. Chu, Laser Cooling of
  Atoms, Ions, or Molecules by Coherent Scattering. \textit{Phys. Rev. Lett.}
  {\bf 84}, 3787-3790 (2000).
  
\bibitem{Kuhn02} A. Kuhn, M. Hennrich, and G. Rempe, Deterministic
  Single-Photon Source for Distributed Quantum Networking. \textit{Phys. Rev.
    Lett.} {\bf 89}, 067901 (2002).
  
\bibitem{Monroe02} for a review, see e.g., C. Monroe, Quantum information
  processing with atoms and photons. \textit{Nature} {\bf 416}, 238-246 (2002).

\bibitem{Horak2001} P. Horak, and H. Ritsch, Dissipative dynamics of Bose
  condensates in optical cavities. \textit{Phys. Rev. A} {\bf 63}, 023603
  (2001).

\bibitem{MabuchiOL96} H.~Mabuchi, Q.A.~Turchette, M.S.~Chapman, and
  H.J.~Kimble, Real-time detection of individual atoms falling through a
  high-finesse optical cavity.  \textit{Opt.~Lett.}~{\bf 21}, 1393-1395 (1996).
  
\bibitem{MunstermannOptCom99} P.~M\"unstermann, T.~Fischer, P.W.H.~Pinkse, and
  G.~Rempe, Single slow atoms from an atomic fountain observed in a
  high-finesse optical cavity. \textit{Opt.~Commun.} {\bf 159}, 63-67 (1999).
  
\bibitem{Sauer03} J.A.~Sauer, K.M.~Fortier, M.S.~Chang, C.D.~Hamley, and
  M.S.~Chapman, Cavity QED with optically transported atoms.
  http://arXiv.org/quant-ph/0309052 (2003).

\bibitem{HechenblaiknerPRA98} G.~Hechenblaikner, M.~Gangl, P.~Horak, and
  H.~Ritsch, Cooling an atom in a weakly driven high-Q cavity.
  \textit{Phys.~Rev.~A} {\bf 58}, 3030-3042 (1998).
  
\bibitem{ChanPRL03} H.W.~Chan, A.T.~Black, and V.~Vuleti{$\rm \acute{c}$},
  Observation of Collective-Emission-Induced Cooling of Atoms in an Optical
  Cavity. \textit{Phys.~Rev.~Lett.}~{\bf 90}, 063003 (2003).
  
\bibitem{Nargorny03} B.~Nagorny, Th.~Els\"asser, and A.~Hemmerich, Collective
  Atomic Motion in an Optical Lattice Formed Inside a High Finesse Cavity.
  \textit{Phys.~Rev.~Lett.}~{\bf 91}, 153003 (2003).
  
\bibitem{Kruse03} D.~Kruse, C. von Cube, C.~Zimmermann, and Ph.W. Courteille,
  Observation of Lasing Mediated by Collective Atomic Recoil.
  http://arXiv.org/quant-ph/0305033 (2003).
  
\bibitem{vanEnkPRA01} S.J. van Enk, J. McKeever, H.J. Kimble, and J. Ye,
  Cooling of a single atom in an optical trap inside a resonator.
  \textit{Phys. Rev. A} {\bf 64}, 013407 (2001).
  
\bibitem{HoodScience00} C.J.~Hood, T.W.~Lynn, A.C.~Doherty, A.S.~Parkins, and
  H.J.~Kimble, The Atom-Cavity Microscope: Single Atoms Bound in Orbit by
  Single Photons. \textit{Science} {\bf 287}, 1447-1453 (2000).
  
\bibitem{PinkseNature00} P.W.H.~Pinkse, T.~Fischer, P.~Maunz, and G.~Rempe,
  Trapping an atom with single photons. \textit{Nature} {\bf 404}, 365-368 (2000).
  
\bibitem{MunstermannPRL99} P.~M\"unstermann, T.~Fischer, P.~Maunz,
  P.W.H.~Pinkse, and G.~Rempe, Dynamics of Single-Atom Motion Observed in a
  High-Finesse Cavity. \textit{Phys.~Rev.~Lett.}~{\bf 82}, 3791-3794 (1999).
  
\bibitem{FischerPRL02} T. Fischer, P. Maunz, P.W.H. Pinkse, T. Puppe, and G.
  Rempe, Feedback on the motion of a single atom in an optical cavity.
  \textit{Phys.  Rev. Lett.}~{\bf 88}, 163002 (2002).

\bibitem{YePRL99} Y.~Ye, D.W. Vernooy, and H.J. Kimble, Trapping of Single
  Atoms in Cavity QED. \textit{Phys. Rev. Lett.} {\bf 83}, 4987-4990 (1999).
  
\bibitem{McKeeverPRL03} J. McKeever, {\it et al.} 
  State-Insensitive Cooling and Trapping of Single Atoms in an Optical Cavity.
  \textit{Phys.  Rev. Lett.} {\bf 90}, 133602 (2003).
  
\bibitem{SavardPRA97} T.A.~Savard, K.M.~O'Hara, and J.E.~Thomas,
  Laser-noise-induced heating in far-off resonance optical traps.
  \textit{Phys.~Rev.~A} {\bf 56}, R1095-R1098 (1997).

\bibitem{Aspect86} \new{A. Aspect, J. Dalibard, A. Heidmann, C. Salomon, and
    C. Cohen-Tannoudji, Cooling Atoms with Stimulated Emission.  \textit{Phys.
      Rev. Lett.} {\bf 57}, 1688-1691 (1986).}
 
\bibitem{Dalibard85} J. Dalibard, and C. Cohen-Tannoudji, Dressed-atom
  approach to atomic motion in laser light: the dipole force revisited.
  \textit{J. Opt. Soc. Am. B} {\bf 2}, 1707-1720 (1985).
  
\bibitem{Griessner03} \new{A. Griessner, D. Jaksch, and P. Zoller, Cavity Assisted
  Nondestructive Laser Cooling of Atomic
  Qubits. http://arXiv.org/quant-ph/0311054 (2003).}
  
\end{thebibliography}
\end{document}